\documentclass[prb,floatfix,twocolumn,showpacs,amsmath,amssymb,superscriptaddress]{revtex4-2}

\usepackage{graphicx}
\usepackage{physics}
\usepackage{ifthen}
\usepackage{color}
\usepackage{bm}
\usepackage{braket}
\usepackage{multirow}
\usepackage{threeparttable} 
\usepackage{hyperref}
\usepackage{xltabular}
\hypersetup{ backref=true,
 pdfnewwindow=true, colorlinks=true,
 linkcolor=blue, anchorcolor=blue,
 citecolor=blue, filecolor=blue,
 menucolor=blue, urlcolor=blue}

\newcommand\mcc[1]{\multicolumn{1}{c}{#1}}
\usepackage{siunitx}
\usepackage{etoolbox}
\robustify\tnote

\begin{document}

\title{Electron-phonon coupling in ferromagnetic Fe--Co alloys from first principles}

\author{Kevin Moseni}
\affiliation{Materials Science and Engineering, University of California Riverside, Riverside, CA
  92521, USA}
\author{Richard B. Wilson}
\author{Sinisa Coh}
\affiliation{Materials Science and Engineering, University of California Riverside, Riverside, CA
  92521, USA}
\affiliation{Mechanical Engineering, University of California Riverside, Riverside, CA
  92521, USA}
  
\begin{abstract}
The measured magnetization dynamics of ferromagnetic iron--cobalt Fe$_{1-x}$Co$_x$ alloys show a strong dependence on the alloy composition, especially near $x=0.25$. Here, we calculate from first principles the electron-phonon coupling strength in Fe$_{1-x}$Co$_x$ alloys for compositions ranging from $x=0$ to $x=0.75$. We find a strong, spin-dependent variation of the electron-phonon coupling strength with alloy composition, with a minimum near $x=0.25$.  We analyze the variation of the electron-phonon interaction with composition, as a function of electron spin, density of states, electron-phonon matrix elements, character of electron wavefunction at the Fermi level, orbital-resolved strength of the phonon perturbing potential, and phonon frequencies. We calculate the electron-phonon energy transfer coefficients, and find that they are in qualitative agreement with the phenomenological electron-phonon energy transfer coefficient deduced from magnetization dynamics experiments. Our findings show that variations in the composition of ferromagnetic alloys can significantly alter the magnetization dynamics and transport properties.
\end{abstract}

\maketitle

\section{Introduction}\label{sec:intro}  
Electron-phonon coupling plays a key role in our understanding of the dynamic properties of ferromagnets. Properties like Gilbert damping,\cite{Gilbert} ultrafast demagnetization,\cite{ultrafast_nickel} and all-optical switching\cite{all_optical} are governed by the electronic bands at the Fermi level and the strength of coupling between the total electron population with phonons. Other properties, such as the spin-dependent Peltier effect\cite{spin_peltier}, giant magneto-thermal effect~\cite{spin_heat_acc,GMTR_nat,spin_transport0,spin_transport1}, and spin-dependent Seebeck effect\cite{seebeck_eph}, are due to differences in majority-spin and minority-spin electronic bands at the Fermi level as well as the different strengths of coupling of majority-spin and minority-spin electron populations with phonons. Engineering these magnetic properties requires an understanding of the material parameters, like electron density of states, electron-phonon matrix elements, and their spin-resolved counterparts, that affect the strength of coupling between electrons and phonons.  Previous work on modeling these spin-dependent properties, such as demagnetization studied in Refs.~\cite{demag0,demag1,demag2}, relies on using heuristic values of the electron-phonon energy transfer coefficients.  Clearly, there is a need to compute these parameters from first-principles in order to obtain a quantitative prediction of the dynamic properties of ferromagnets.

Previous first-principles studies of the electron-phonon coupling strength in magnets have focused on only a small subset of ferromagnetic materials, namely elemental ferromagnets Ni, Fe, and Co. However, many ferromagnets are alloys of transition metals. One such example is the Fe--Co alloy. Magnetic damping in the Fe--Co alloy was only recently measured.\cite{schoen2016} Interestingly, Ref.~\onlinecite{schoen2016} reports significant changes in damping as a function of the Fe--Co alloy composition. In particular, the lowest damping was measured in the Fe$_{0.75}$Co$_{0.25}$ alloy, the same alloy composition that the density of states is lowest. Measurements of minimal damping in this composition have been reported by others.\cite{Lee2017,Weber_2019} More recent experiments showed that the demagnetization response is maximized in this composition due to a minimum in the electron-phonon energy transfer coefficient,\cite{Ramya2021} a quantity that depends not only on the density of states but also the electron-phonon matrix elements. 

Our goal in this work is to understand the composition dependence of the electron-phonon coupling strength in the ferromagnetic Fe--Co alloy. As we detail in Sec.~\ref{sec:elph_alloys}, surprisingly few computational first-principles studies of the electron-phonon coupling strength in ferromagnetic alloys exist, aside from the study of the electron-phonon energy transfer coefficient in Ni-based alloys.\cite{Nialloys} Although $\lambda$ has been reported for pure Fe and Co,~\cite{verstraete2013} no results have been reported for $\lambda$ in Fe--Co alloys. On the other hand, similar studies exist for many non-magnetic alloys as a function of their composition. The electron-phonon coupling strength in some alloys, such as V--Cr, varies linearly with composition,\cite{MCMILLAN} as one might expect from a rule of mixtures. But in other cases, such as Nb--Mo, there is non-linear dependence of the electron-phonon coupling strength on the composition.\cite{NbMo_exp} 

In this work we use first-principles density functional theory to compute the electron-phonon coupling strength $\lambda$ in Fe--Co alloys. We find that $\lambda$ varies strongly, and non-monotonically, with alloy composition, reaching a minimum near the aforementioned Fe$_{0.75}$Co$_{0.25}$ composition. We perform a spin decomposition of $\lambda$ and find that the majority-spin contribution to $\lambda$ monotonically decreases with the alloy composition, while the opposite is true for minority-spin. We further analyze the quantities driving the composition dependence of $\lambda$ and find that while the density of states plays a dominant role, the magnitude of the electron-phonon matrix elements also varies strongly with the alloy composition.  We compare our theoretical results with experimental estimates of the electron-phonon energy transfer coefficient in Fe--Co alloys across a range of compositions~\cite{Ramya2021} and find that our results are qualitatively in agreement with the experiment.

The structure of this paper is as follows. In Sec.~\ref{sec:elph_alloys}, we discuss previous experimental and theoretical values of the electron-phonon coupling strength in metallic alloys. In Sec.~\ref{sec:methods}, we detail how we decompose the electron-phonon interaction strength into their spin components.  Then in Sec.~\ref{sec:RandA}, we show the results of the first-principles density functional theory for the electron-phonon coupling strength in ferromagnetic Fe--Co alloys. We analyze separately the strengths of majority- and minority-spin components in order to reveal the origin of the nonmonotonicity. We also compare our results with the experiments. 

\section{Electron phonon coupling in alloys}\label{sec:elph_alloys}

\begin{figure*}
\includegraphics[width = 6.1 in]{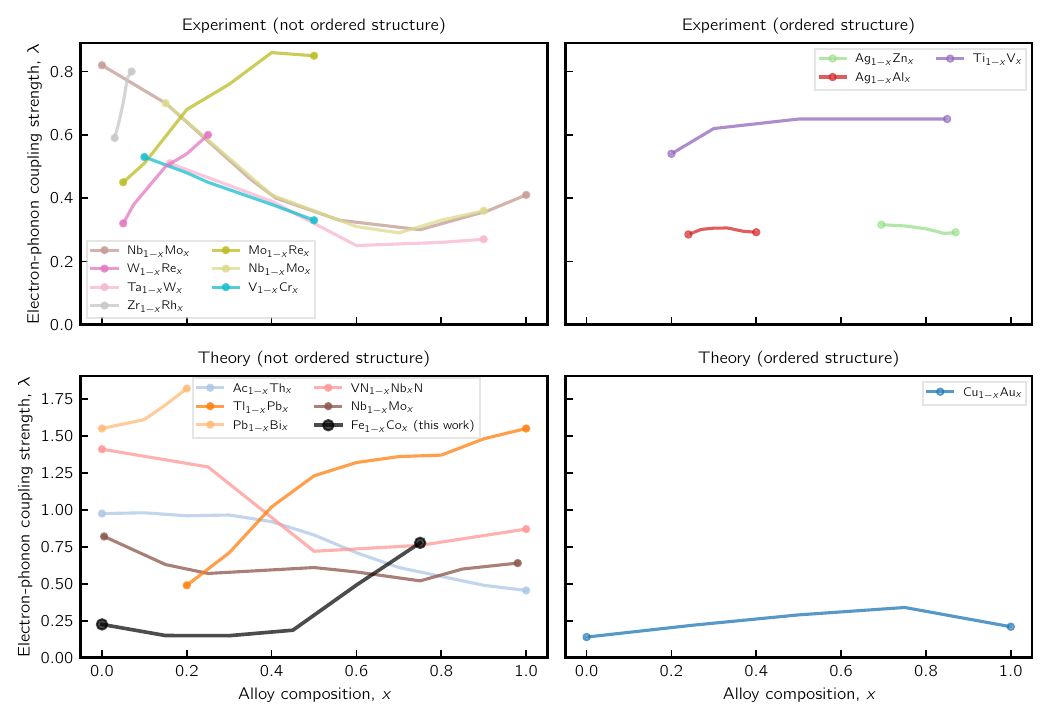}
\caption{\label{fig:2by2} Dependence of the electron-phonon coupling strength ($\lambda$) on composition ($x$). Experimental results are from Refs.~\onlinecite{MCMILLAN,NbMo_exp,AgZn_exp,AgAl_exp}. Theoretical results are from Refs.~\onlinecite{AcTh,TlPbBi,VNNbN,NbMo_th,CuAu}.}
\end{figure*}

The dependence of $\lambda$ on alloy composition has been studied for various nonmagnetic alloy systems. These include empirical values of $\lambda$, estimated from the ratio of the Debye temperature and the superconducting transition temperature, in ordered and disordered alloys,\cite{cheng1962,MCMILLAN} as well as first principles calculated $\lambda$ in virtual crystal approximated alloys\cite{donato1982fermi,AcTh,TlPbBi}, ordered alloys,\cite{CuAu} and disordered alloys.\cite{NbMo_th,VNNbN} Naively, one might think that $\lambda$ in an alloy A$_{1-x}$B$_x$ with composition $x$ could be calculated by linear interpolation from the electron-phonon coupling strength of metal A ($\lambda_{\rm A}$) to that of metal B ($\lambda_{\rm B}$). To evaluate how well the linear model approximates $\lambda$, we computed 
\begin{equation}\label{eq:lambda_linear}
\lambda^{\rm linear}= (1-x) \lambda_{\rm A} + x \lambda_{\rm B}
\end{equation}
for the various alloys in the literature. For many of the alloys, the literature contains values for $\lambda$ for only part of the range of compositions. For these, we linearly interpolate from $\lambda$ at the lowest available composition $x_{\rm min}$ to $\lambda$ at the highest available composition $x_{\rm max}$ using a simple generalization of Eq.~\ref{eq:lambda_linear}, 
\begin{equation}\label{eq:lambda_linear_gen}
\lambda^{\rm linear} = 
\frac{x_{\rm max}-x}{x_{\rm max}-x_{\rm min}} 
\lambda_{x_{\rm min}}
+ 
\frac{x-x_{\rm min}}{x_{\rm max}-x_{\rm min}}
\lambda_{x_{\rm max}}.
\end{equation}
However, as expected, this naive linear approximation is valid only for a few alloys.  The deviations from the linear behavior are quite commonly reported in the literature,\cite{MCMILLAN,NbMo_exp,AgZn_exp,AgAl_exp,cheng1962, AcTh, TlPbBi, VNNbN, NbMo_th, CuAu} as summarized in Fig.~\ref{fig:2by2}. The vertical axis in Fig.~\ref{fig:2by2} shows $\lambda$ and the horizontal scale of the figure shows the composition ranging from $x_{\rm min}$ to $x_{\rm max}$ for each alloy based on available data. 

\begin{figure*}
\includegraphics[width = 6.1 in]{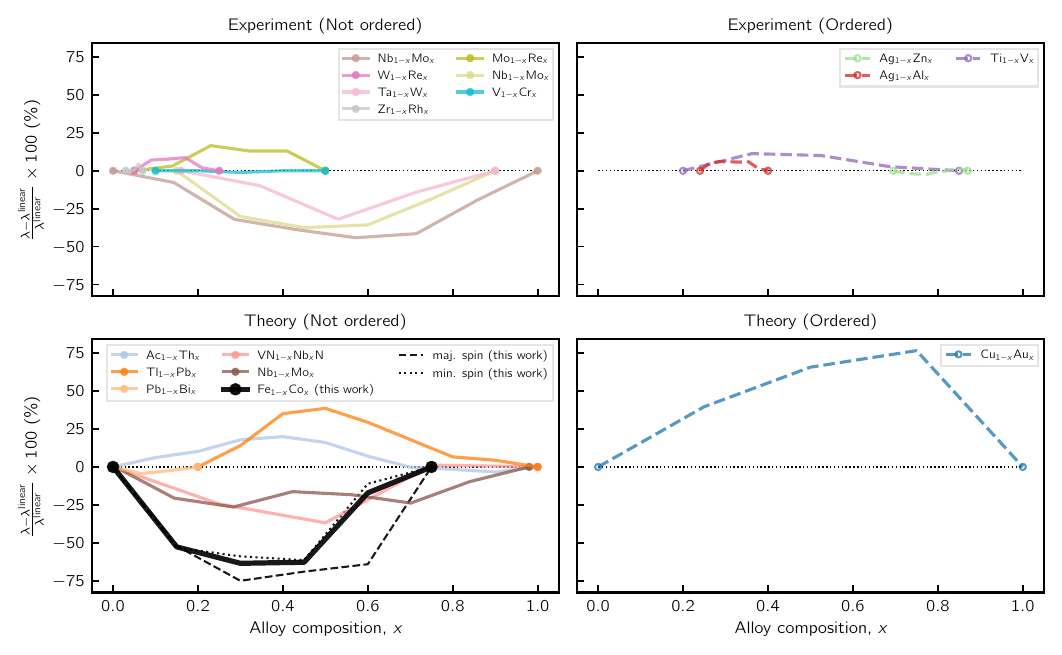}
\caption{\label{fig:2by2_deviations} Same as Fig.~\ref{fig:2by2} but now instead of $\lambda$ we show deviation of $\lambda$ from the linear model given by Eq.~\ref{eq:lambda_linear_gen}.}
\end{figure*}

We further quantify the deviation from the naive linear model by computing $\frac{\lambda-\lambda^{\rm linear}}{\lambda^{\rm linear}}$ for each alloy, as shown in Fig.~\ref{fig:2by2_deviations}. 

As can be seen from Fig.~\ref{fig:2by2_deviations}, the V$_{1-x}$Cr$_{x}$ alloy in the range of $x=0.1$ to $x=0.5$ has a reported relative deviation from the naive linear estimation of only -1\%. On the other hand, in Ti--V and Mo--Re alloys, the deviation from linear dependence is up to 20\%, while in the Pb--Tl, VN--NbN, Nb--Mo, and Ta--W alloys it is even greater, around $\pm$ 40\%. The largest deviation from the linear regime, about 80\%, was found for ordered Cu$_{3/4}$Au$_{1/4}$ alloy.  We note that the linear dependence of $\lambda$ on composition in some alloy systems (Zr--Rh, Ag--Zn, Pb--Bi, and Ag--Al) is due to a relatively narrow reported range of composition ($x$).

Our results on the ferromagnetic Fe--Co alloy are shown in the bottom left panels of Fig.~\ref{fig:2by2} and Fig.~\ref{fig:2by2_deviations} in black. We find that $\lambda$ of Fe$_{0.25}$Co$_{0.75}$ is about 70\% smaller than that predicted by the naive linear model, $\lambda^{\rm linear}$.

We now turn our attention to the definitions of key quantities that we will use to study the electron-phonon coupling in Fe--Co alloys. 


\section{Methods}\label{sec:methods}

The strength of electron-phonon coupling is often measured by a single dimensionless number, $\lambda$.\cite{MCMILLAN, GIUSTINO2017} The coupling constant $\lambda$ is defined as a double sum over the Fermi surface,
\begin{align}\label{eq:lambda}
\lambda = & ~\frac{1}{N_{\rm F}}\frac{1}{N^{ }_{k} N^{ }_{q}} \sum_{q\nu} \frac{1}{\hbar \omega_{q\nu}} \sum_{mn,k}   |g_{m n \nu}(k,q)|^2 \notag \\
          & ~\times \delta(\epsilon_{nk}-\epsilon_{\rm F})  \delta(\epsilon_{m k+q}-\epsilon_{\rm F}).
\end{align} 
We denote the total density of states at the Fermi level, for both spin channels, as $N_{\rm F}$, the number of sampled $k$-points in the electron Brillouin zone as $N_k$, and number of sampled $q$-points in the phonon Brillouin zone as $N_q$. The phonon frequency is $\omega_{q\nu}$. The first sum in Eq.~\ref{eq:lambda} goes over phonon $q$ points and branches $\nu$, while the second sum goes over $k$ points and electronic bands $m$ and $n$.  For now, spin indices are incorporated into the band indices. The energy conservation is maintained by the two delta functions. The electron-phonon matrix element $g_{m n \nu}(k,q)$ from Eq.~\ref{eq:lambda} is defined as
\begin{equation}\label{eq:mat_elems}
g_{m n \nu}(k,q) = \sqrt{\frac{\hbar}{2m\omega_{q\nu}}} \langle \psi_{m k+q} | \partial_{q\nu} v | \psi_{nk} \rangle.
\end{equation}
Here, $\partial_{q\nu} v$ is the periodic modulation of the crystal potential $v$ due to phonon $(q,\nu)$ that allows an electron in the state $(n,k)$ to scatter to state $(m, k+q)$. 

\subsection{Spin-resolved measures of electron-phonon interaction}\label{subsec:spin_quantities}

For materials with a small spin-orbit interaction, the electronic states have a well-defined spin.  Therefore, from now on we use $\sigma = \ \uparrow$ or $\sigma = \ \downarrow$ to index the Bloch state $\psi^\sigma_{nk}$ with a specific spin state.  In the small spin-orbit limit the dominant scattering is between states of equal spin,
\begin{equation}\label{eq:g_spin}
g^{\sigma}_{m n \nu}(k,q) = \sqrt{\frac{\hbar}{2m\omega_{q\nu}}} \langle \psi^\sigma_{m k+q} | \partial_{q\nu} v | \psi^\sigma_{nk} \rangle.
\end{equation}
Under this assumption, the sums over $m$ and $n$ in Eq.~\ref{eq:lambda} trivially decompose into two separate sums, one where both bra and ket states have $\sigma = \ \uparrow$ and another where both states have $\sigma = \ \downarrow$ component.   

With this separation, $\lambda$ from Eq.~\ref{eq:lambda} decomposes into two terms, one for each spin channel $\sigma$,
\begin{align}
\lambda=\lambda^\uparrow + \lambda^\downarrow.
\label{eq:additive_lambda}
\end{align} 
The spin-dependent quantity $\lambda^\sigma$ is defined as
\begin{align}\label{eq:lambda_spin}
\lambda^{\sigma} = & \frac{1}{N_{\rm F}^{\uparrow} + N_{\rm F}^{\downarrow}}\frac{1}{N_{k} N_{q}} 
                   \sum_{q\nu} \frac{1}{\hbar \omega_{q\nu}}          
                  \sum_{m n k}  |g^{\sigma}_{m n \nu}(k,q)|^2 \notag \\
                & ~\times \delta \left( \epsilon_{nk}^{\sigma}-\epsilon_{\rm F} \right) \delta\left( \epsilon_{m k+q}^{\sigma} -\epsilon_{\rm F} \right).
\end{align}
The only difference between total $\lambda$ from Eq.~\ref{eq:lambda} and spin-resolved $\lambda^{\sigma}$ from Eq.~\ref{eq:lambda_spin} is that the former involves a sum over all states, while the latter sums over states with fixed spin $\sigma$.  Importantly, in Eq.~\ref{eq:lambda_spin} of the spin-resolved quantity $\lambda^{\sigma}$ we still have the total density of states $N_{\rm F} = N_{\rm F}^{\uparrow} + N_{\rm F}^{\downarrow}$ in the denominator.

In what follows it will be convenient to work with the Eliashberg spectral function, $\alpha^2 F (\omega)$, as it keeps track of the phonon frequencies $\omega$ which connect the electron states with wavevector $k$ to a state with the wavevector $k+q$, both of which are at or near the Fermi surface.\cite{eliashberg1960OG,MCMILLAN} We denote the wavevector of a phonon with $q$.   In analogy to the decomposition of $\lambda$ into two spin components, we decompose the Eliashberg spectral function into contributions from each spin channel, $\alpha^2 F^{\sigma} (\omega)$.  We define this spin-resolved quantity so that integrating over $d \omega / \omega$ gives back our additive spin-resolved $\lambda^{\sigma}$,
\begin{equation}\label{eq:lam_a2f}
\lambda^{\sigma} = 2 \int_0^{\infty} \frac{d\omega}{\omega} \alpha^2 F^\sigma (\omega).
\end{equation}
Therefore, clearly, in analogy to Eq.~\ref{eq:additive_lambda} we have
\begin{align}\label{eq:additive_a2f}
\alpha^2F(\omega) = \alpha^2F^\uparrow(\omega) + \alpha^2F^\downarrow(\omega)
\end{align}
where $\alpha^2F(\omega)$ is the conventional Eliashberg spectral function used, for example, in the study of superconductivity and elsewhere.

\subsubsection{Alternative definition: non-additive $\lambda$}

Previous work on electron-phonon coupling in ferromagnetic metals, Ref.~\onlinecite{verstraete2013}, introduced a different measure of spin-resolved electron-phonon coupling strength.  Instead of dividing by the total density of states, as in Eq.~\ref{eq:lambda_spin}, one divides by a spin-resolved density of states.  This leads to a quantity $\lambda^{\sigma}_{\rm nadd}$ which is related to $\lambda^{\sigma}$ simply as,
\begin{align} \label{eq:lambda_nadd}
\lambda^{\sigma}_{\rm nadd}
=
\frac{N_{\rm F}^{\uparrow} + N_{\rm F}^{\downarrow}}{N_{\rm F}^{\sigma}} 
\lambda^{\sigma}.
\end{align}
Clearly, such $\lambda^{\sigma}_{\rm nadd}$ is no longer additive over spins.  Instead, using Eqs.~\ref{eq:additive_lambda} and \ref{eq:lambda_nadd} we have the following relation,
\begin{align} \label{eq:lambda_from_lambda_nadd}
\lambda =
\frac{N_{\rm F}^\uparrow
\lambda^\uparrow_{\rm nadd} 
+
N_{\rm F}^\downarrow
\lambda^\downarrow_{\rm nadd}
}{N_{\rm F}^{\uparrow} + N_{\rm F}^{\downarrow}} .
\end{align} 
While definition of $\lambda^\sigma_{\rm nadd}$ somewhat complicates its relation to the total $\lambda$, a common measure of electron-phonon interaction in the literature, the quantity $\lambda^{\sigma}_{\rm nadd}$ has another convenient property. If we for a moment neglect some of the details of the electronic band structure and assume that the electron-phonon matrix elements on the Fermi surface are independent of band indices $m$ and $n$, then the matrix element can be taken out of the double sum, and the sum over the two delta functions gives us $\left(N_{\rm F}^{\sigma}\right)^2$ which partially cancels $N_{\rm F}^{\sigma}$ from the denominator.  Therefore, relationship between non-additive $\lambda^{\sigma}_{\rm nadd}$  and the average electron-phonon matrix element $\langle g_{\sigma}^2 \rangle$ for a given spin-channel is particularly straightforward,
\begin{align} \label{eq:average_g}
\lambda^{\sigma}_{\rm nadd} = N_{\rm F}^{\sigma} \langle g_{\sigma}^2 \rangle.
\end{align}
On the other hand, if one wanted to relate our additive $\lambda^{\sigma}$ to the average matrix element, combining Eqs.~\ref{eq:lambda_nadd} and \ref{eq:average_g} would lead to a less intuitive relationship with the average matrix element strength,
\begin{equation}\label{eq:lam_simple_spin}
\lambda^{\sigma} = \frac{(N^\sigma_{\rm F})^2}{N_{\rm F}^{\uparrow} + N_{\rm F}^{\downarrow}} \langle g_\sigma^2 \rangle.
\end{equation}

\subsection{Relationship to electron-phonon energy transfer coefficient $G$}\label{sec:measurable}

Now we relate our $\lambda$ and $\lambda^\sigma$ to the electron-phonon energy transfer coefficient $G$.  This coefficient describes the rate of energy transfer between electrons and phonons per unit volume and per temperature difference of electrons and phonons. Within the two-temperature model of electrons and phonons for a nonmagnetic system,\cite{Allen_gep} $G$ is proportional to the total density of states $N_{\rm F} = N_{\rm F}^{\uparrow} + N_{\rm F}^{\downarrow}$ and $\lambda$. This model implicitly assumes that majority and minority spins are thermalized with each other at all times.  With this assumption, the effective $G$ for a spin-polarized system is given as,
\begin{align}
G = \frac{\pi  k_{B} \hbar}{ V_{\rm c}}\left[\left( N_{\rm F}^\uparrow +  N_{\rm F}^\downarrow \right) \left( \lambda^\uparrow + \lambda^\downarrow \right) \right] \langle \omega^2 \rangle
\label{eq:G1} 
\end{align}
using additive $\lambda^{\sigma}$, or equivalently 
\begin{align}
G =  \frac{\pi  k_{B} \hbar}{ V_{\rm c}} \left(N_{\rm F}^\uparrow \lambda_{\rm nadd}^\uparrow + N_{\rm F}^\downarrow \lambda_{\rm nadd}^\downarrow \right) \langle\omega^2\rangle
\label{eq:G2}
\end{align}
using the non-additive variant. Here $V_{\rm c}$ is the unit cell volume (per atom), and $\langle\omega^2\rangle $ is a weighted average of the square of the phonon frequency, defined as\cite{MCMILLAN} 
\begin{align}\label{eq:omega2}
\langle\omega^2\rangle = \frac{2}{\lambda^\uparrow + \lambda^\downarrow} \int_0^\infty d\omega~\omega \left[\alpha^2F^\uparrow(\omega) + \alpha^2F^\downarrow(\omega)\right],
\end{align}
with units of meV$^2$.
Equations~\ref{eq:G1} and \ref{eq:G2} give a numerically equal energy transfer coefficient $G$.  Therefore, additive $\lambda^\sigma$ or non-additive $\lambda_{\rm nadd}^\sigma$ can be used to compute $G$. The only difference between the two equations is that they implicitly assume a different thermodynamic description of the electronic system.  In the case of Eq.~\ref{eq:G1} one considers the electronic system in a ferromagnet as one unified thermal reservoir.  The effective electron-phonon coupling strength for this system is then the sum of $\lambda^{\uparrow}$ and $\lambda^{\downarrow}$.  In contrast, in the case of Eq.~\ref{eq:G2}, one imagines the electronic system as consisting of two reservoirs, one for each spin, and each subsystem has its density of states and effective electron-phonon coupling strength.

As with $\lambda$, we can separate $G$ into its majority-spin part $G^\uparrow$ and minority-spin part $G^\downarrow$,
\begin{align}
G = G^\uparrow + G^\downarrow
\end{align}
where the electron-phonon energy transfer coefficient for spin channel $\sigma$ is defined as 
\begin{align}
G^\sigma = \frac{\pi k_{B} \hbar}{ V_{\rm c}} \left( N_{\rm F}^{\uparrow} + N_{\rm F}^{\downarrow} \right) \ 2 \int_0^\infty d\omega~\omega \alpha^2F^\sigma(\omega).
\label{eq:G_spin}
\end{align}

\subsection{Details of density functional theory calculation}\label{subsec:details}

In this work, we use the \textsc{quantum espresso} computer package\cite{QE_2009, QE_2017} for density functional theory calculations. We use the GGA-PBE exchange-correlation functional.~\cite{PBE} We do not include Hubbard $U$ correction, as it leads to an overestimation of both the lattice parameter and magnetic moment of Fe.\cite{Iron_plusU} We choose the non-relativistic ONCV pseudopotentials\cite{ONCV_2013} for Fe and Co. We use 90~Ry kinetic-energy cutoff and $18^3$~$k$~points on a uniform grid to converge wavefunctions. We use the density functional perturbation theory to calculate the phonons on a coarse grid of $4^3$~$q$~points. We use Wannier90\cite{marzari1997maximally, souza2001maximally, W90updated_2014} to construct Wannier functions with $sp^3d^2$-like and $t_{2 \rm g}$-like characters. We set the frozen window to range from 30 eV below to 5 eV above the Fermi level.

We compute the electron-phonon matrix elements with EPW.\cite{GIUSTINO2007, ponce2016epw} The convergence of $\lambda^\sigma$ requires a small smearing parameter and a large number of $k$-points and $q$-points. We vary these parameters until convergence in the limit of zero smearing and infinitely many $k$ and $q$ points.  We determine that a smearing of 0.01 eV, $72^3$~$k$~points, and $24^3$~$q$~points are sufficient to converge $\lambda^\sigma$. To converge the density of states at the Fermi energy, we use $192^3$~$k$~points. We modified the native EPW package to support spin-polarized calculations. All of our calculations are done without including the spin-orbit interaction, as it is relatively weak in 3$d$ metals such as Fe and Co.

\subsection{Virtual-crystal approximation}\label{subsec:VCA}
\begin{figure}
\includegraphics{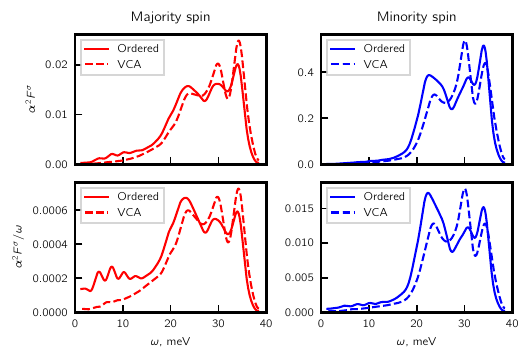}
\caption{\label{fig:vca_v_supercell} $\alpha^2F^{\sigma}$ and $\alpha^2F^{\sigma}/\omega$ for FeCo alloy at $x = 0.5$ calculated for an ordered supercell (solid lines) and within the VCA (dashed lines) for majority spins (left column) and minority spins (right column).}
\end{figure}

In this work, we focus on Fe$_{1-x}$Co$_x$ alloys in the range of concentrations $x$ from 0 to 0.75. Pure iron ($x=0$) at room temperature adopts a body-centered cubic structure (bcc, space group $Im\bar{3}m$). When Co is added to Fe, the structure remains bcc, but Fe and Co atoms are randomly arranged on a bcc lattice. At a higher concentration of Co ($x=0.5$) there is a ordering of Fe and Co atoms (B$_2$ phase, space group $Pm\bar{3}m$).\cite{nishizawa1984,matsuda2012development} At an even higher concentration of Co (above around $x=0.75$) the bcc phase is no longer favored and the preferred phase is hexagonal close-packed (hcp). 
 
We model Fe$_{1-x}$Co$_x$ alloys within the virtual crystal approximation (VCA).\cite{VCA_1931,VCA_2000}  Briefly, we use two elemental pseudopotentials and combine them with the desired fraction of each element to create an effective pseudopotential of the alloy. Then we place the pseudopotential of the alloy in the primitive cell of the bcc lattice (2a site of space group $Im\bar{3}m$) to generate the bcc alloy. The VCA approach offers a computational advantage, because we can use a single atom basis to represent arbitrary alloy compositions $x$, eliminating the need for computationally expensive supercell calculations. Nevertheless, we computed $\lambda$ at $x=0.5$ using both approaches. To construct the supercell, we place Fe in the 1a site and Co in the 1b site of the space group $Pm\bar{3}m$, resulting in an ordered (B$_2$) Fe$_{1/2}$Co$_{1/2}$ alloy. 

On converging both calculations we find that in the ordered Fe$_{1/2}$Co$_{1/2}$ alloy, $\lambda$ is 0.496 while in the VCA approach $\lambda$ is surprisingly similar, 0.491. We find a similar agreement in $\lambda^\uparrow$ and $\lambda^\downarrow$. To further compare these calculations, we first recall that by Eq.~\ref{eq:lam_simple_spin} our $\lambda^\sigma$ is the product of $\frac{(N^{\sigma}_{\rm F})^2}{N^\uparrow_{\rm F}+N^\downarrow_{\rm F}}$ and the average matrix element $\langle g_{\sigma}^2 \rangle$. Comparing the values obtained from both methods, we find that the VCA overestimates $\frac{(N^{\uparrow}_{\rm F})^2}{N^\uparrow_{\rm F}+N^\downarrow_{\rm F}}$ by 20\%, and underestimates $\langle g_{\uparrow}^2 \rangle$ by 10\%. The opposite is true for the minority spins: underestimation of $\frac{(N^{\downarrow}_{\rm F})^2}{N^\uparrow_{\rm F}+N^\downarrow_{\rm F}}$ and overestimation of $\langle g_{\downarrow}^2 \rangle$. Therefore, the remarkably close agreement in $\lambda^\sigma$ and $\lambda$ obtained from VCA and supercell is partly due to accidental partial cancellation of errors in the densities of states and electron-phonon matrix elements.  To further analyze our results, we compare the spin-resolved Eliashberg spectral functions (shown in Fig.~\ref{fig:vca_v_supercell}) and we find that the supercell approach gives a spectral function that has more features, especially at low frequencies. Furthermore, we find some weight redistribution among the peaks at the higher part of the phonon spectrum. We assign these differences to the folding of the band structure in the supercell approach that is absent in the VCA. Given the large number of $k$ and $q$ points needed to converge these calculations, computing $\lambda$ for other values of $x$ with supercells, or taking into account disorder with special quasi-random structures\cite{SQS_Zunger}, quickly becomes computationally prohibitive.

\section{Results \& Discussion}\label{sec:RandA}

\begin{figure}[htp]
\includegraphics{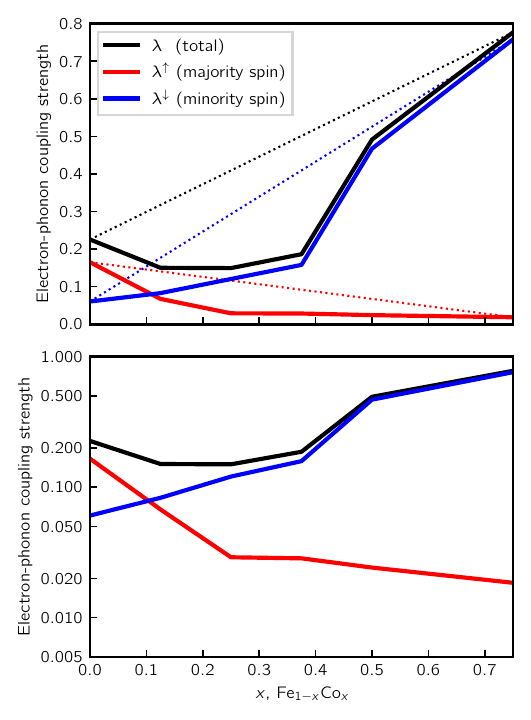}
\caption{\label{fig:lambda} Electron-phonon coupling strength $\lambda$ (black line) is a non-monotonic function of the alloy composition $x$. The majority-spin $\lambda^{\uparrow}$ (red line) decreases with composition as the minority-spin $\lambda^{\downarrow}$ (blue line) increases. The top panel has a linear vertical scale, and the bottom panel has a logarithmic vertical scale. The results shown here are from VCA.}
\end{figure}

\begin{figure}[htp]
\includegraphics{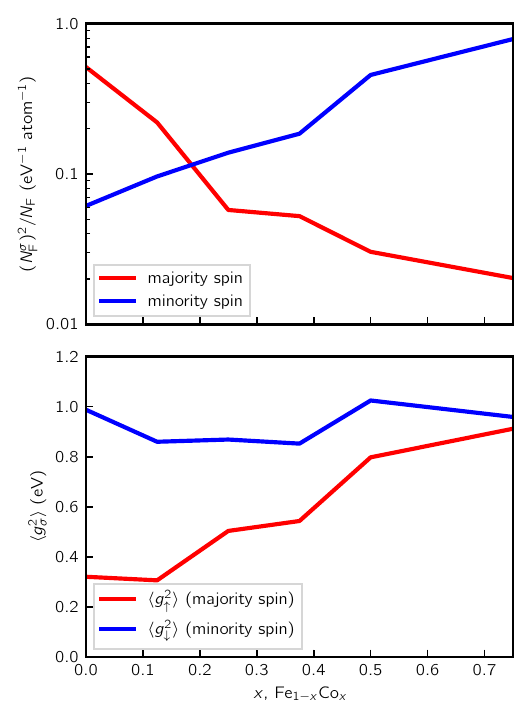}
\caption{\label{fig:expl2} Compositional dependence of $\frac{\left(N^{\sigma}_{\rm F} \right)^2}{N_{\rm F}^{\uparrow} + N_{\rm F}^{\downarrow}}$ (top panel, log scale) and average electron-phonon matrix element $\langle g_{\sigma}^2 \rangle$ (bottom panel, linear scale). With increasing $x$ the majority spin $\frac{\left(N^{\uparrow}_{\rm F}\right)^2}{N_{\rm F}^{\uparrow} + N_{\rm F}^{\downarrow} }$ decreases, while the minority spin $\frac{\left(N^{\downarrow}_{\rm F}\right)^2}{N_{\rm F}^{\uparrow} + N_{\rm F}^{\downarrow}}$ increases. The average matrix element of the majority spins $\langle g_{\uparrow}^2 \rangle$ increases with $x$ while the minority spin $\langle g_{\downarrow}^2 \rangle$ is constant.}
\end{figure}

Now we discuss the results of our calculations of $\lambda$ in Fe$_{1-x}$Co$_x$ for $x$ between $0$ and $0.75$ in steps of $0.125$. In the case of pure Fe ($x=0$), we find $\lambda$ to be 0.23. As we increase the Co concentration $x$, we find that $\lambda$ first decreases for small $x$ and then increases substantially for larger $x$, as shown by the black line in Figure~\ref{fig:lambda}.  The numerical values of $\lambda$ are also shown in Table~\ref{tab:results}. Between concentrations $x=0$ and $x=0.125$, $\lambda$ decreases by a factor of 1.5.  As the concentration increases further, we find that $\lambda$ increases by a factor of 5 between $x=0.25$ and $x=0.75$.   Table~\ref{tab:results} also contains numerical values of the electron-phonon energy transfer coefficient $G$ calculated using Eqs.~\ref{eq:G1} and \ref{eq:omega2}.

\begin{table*}[!p]
\caption{\label{tab:results} Quantities related to the electron-phonon coupling in the Fe$_{1-x}$Co$_x$ alloys.  $N^\sigma_{\rm F}$, $N_{\rm F}$ in units of eV$^{-1}$atom$^{-1}$, $\langle g_{\sigma}^2 \rangle$ in units of eV, $G$ in units of 10$^{17}$\,Wm$^{-3}$K$^{-1}$, magnetic moment $\mu$ in units of $\mu_B/{\rm atom}$, and cell volume (per atom) $V_{\rm c}$ in units of \AA$^3$. The theoretical results from Refs.~\onlinecite{Medvedev,Oppeneer,trinastic2013first,verstraete2013,cazzaniga2010ab,petrov2013thermal} and experimental result from Ref.~\onlinecite{ma14112755} are included for comparison.}
\begin{threeparttable}
\begin{ruledtabular}
\begin{tabular}{llSSSSSSS}
& 
& \mcc{Fe ($x=0$)}  
& \mcc{$x=0.125$} 
& \mcc{$x=0.25$} 
& \mcc{$x=0.375$}
& \multicolumn{2}{c}{$x=0.5$}
& \mcc{$x=0.75$} \\
\cline{7-8}
&  & \mcc{Ordered} & \mcc{VCA} & \mcc{VCA} & \mcc{VCA} & \mcc{VCA} & \mcc{Ordered} & \mcc{VCA}  \\
\hline
$N^\uparrow_{\rm F}$ & 
This work & 
0.693 & 0.366 & 0.147 & 0.151 & 0.148 & 0.145 & 0.147  \\
& Ref.~\onlinecite{verstraete2013} & 0.519 & & & & & & \\
& Ref.~\onlinecite{trinastic2013first} & 0.76 & & 0.12 &  & 0.12 & & \\
& Ref.~\onlinecite{cazzaniga2010ab} & 0.70 & & & & & & \\

\cline{2-9}

$N^\downarrow_{\rm F}$ & This work & 
0.239 & 0.242 & 0.228 & 0.284 & 0.574 & 0.634 & 0.918 \\ 
& Ref.~\onlinecite{verstraete2013} & 0.312 & & & & & & \\ 
& Ref.~\onlinecite{trinastic2013first} & 0.24 & & 0.28 & & 0.58 & & \\
& Ref.~\onlinecite{cazzaniga2010ab} & 0.27 & & & & & & \\

\cline{2-9}

$N_{\rm F}$ & This work & 
0.932 &	0.608 &	0.375 &	0.435 &	0.722 &	0.779 &	1.065\\ 
& Ref.~\onlinecite{verstraete2013} & 0.831 & & & & & & \\ 
& Ref.~\onlinecite{trinastic2013first} & 1.00 & & 0.40 & & 0.70 & & \\ 
& Ref.~\onlinecite{cazzaniga2010ab} & 0.970 & & & & & & \\

\cline{2-9}

$\langle g_{\uparrow}^2 \rangle$ & This work & 
0.320 & 0.306 & 0.503 & 0.543 & 0.797 & 0.966 & 0.912 \\ 
& Ref.~\onlinecite{verstraete2013} & 0.131\tnote{a} & & & & & & \\

\cline{2-9}

$\langle g_{\downarrow}^2 \rangle$ & This work & 
0.987 & 0.860 & 0.868 & 0.852 & 1.024 & 0.912 & 0.959 \\  
& Ref.~\onlinecite{verstraete2013} & 0.561\tnote{a} & & & & & & \\

\cline{2-9}

$\lambda_{\rm nadd}^{\uparrow}$ & This work & 
0.222 & 0.112 & 0.074 & 0.082 & 0.118 & 0.140 & 0.134 \\
& Ref.~\onlinecite{verstraete2013} & 0.068 & & & & & & \\  

\cline{2-9}

$\lambda_{\rm nadd}^{\downarrow}$ & This work & 
0.236 & 0.208 & 0.198 & 0.242 & 0.588 & 0.578 & 0.88 \\ 
& Ref.~\onlinecite{verstraete2013} & 0.175 & & & & & & \\ 

\cline{2-9}

$\lambda^{\uparrow}$ & This work &
0.165 & 0.067 & 0.029 & 0.028 & 0.024 & 0.026 & 0.018 \\
& Ref.~\onlinecite{verstraete2013} & 0.042\tnote{b} & & & & & & \\ 

\cline{2-9}

$\lambda^{\downarrow}$ & This work & 
0.061 & 0.083 & 0.120 & 0.158 & 0.467 & 0.470 & 0.759 \\ 
& Ref.~\onlinecite{verstraete2013} & 0.066\tnote{b} & & & & & & \\

\cline{2-9}

$\lambda$ & This work & 
0.226 & 0.150 & 0.149 & 0.186 & 0.491 & 0.496 & 0.777 \\ 
& Ref.~\onlinecite{verstraete2013} & 0.108\tnote{c} & & & & & & \\

\cline{2-9}

$G^{\uparrow}$ & This work & 
4.76 & 1.69 & 0.499 & 0.551 & 0.792 & 0.637 & 0.641 \\ 

\cline{2-9}

$G^{\downarrow}$ & This work & 
2.49 & 2.34 & 2.19 & 3.46 & 15.1 & 14.8 & 27.2 \\

\cline{2-9}

$G$ & This work & 
7.25 & 4.03 & 2.68 & 4.01 & 15.9 & 15.5 & 27.9 \\
& Ref.~\onlinecite{petrov2013thermal} & 7.00 & & & & & \\
& Ref.~\onlinecite{Medvedev} & 20.8 & & & & & \\
& Ref.~\onlinecite{Oppeneer} & 10.5 & & & & & \\
& Ref.~\onlinecite{ma14112755} & 8.80\tnote{d} & & & & &\\
& Ref.~\onlinecite{ma14112755} & 9.40\tnote{e}  & & & & &\\

\cline{2-9}

$\mu$ & This work & 2.25 &  2.37 &  2.34 &  2.35 &  2.23 &  2.28 & 2.01  \\

\cline{2-9}

$V_{\rm c}$ & This work & 11.50 & 11.57 & 11.53 & 11.49 & 11.42 & 23.16  & 11.27 \\
\end{tabular}
\end{ruledtabular}
\begin{tablenotes}\footnotesize
\item[a] Calculated by inserting $\lambda_{\rm nadd}^{\sigma}$ and $N^\sigma_{\rm F}$ from Ref.~\onlinecite{verstraete2013} into our Eq.~\ref{eq:average_g}.
\item[b] Calculated by inserting $\lambda_{\rm nadd}^{\sigma}$ and $N^\sigma_{\rm F}$ from Ref.~\onlinecite{verstraete2013} into our Eq.~\ref{eq:lambda_nadd}.
\item[c] Calculated by inserting $\lambda_{\rm nadd}^{\sigma}$ and $N^\sigma_{\rm F}$ from Ref.~\onlinecite{verstraete2013} into our Eq.~\ref{eq:lambda_from_lambda_nadd}.
\item[d] Direct heating of the sample, as discussed in Ref.~\onlinecite{ma14112755}.
\item[e] Indirect heating of the sample, as discussed in Ref.~\onlinecite{ma14112755}
\end{tablenotes}
\end{threeparttable}
\end{table*}

\subsection{Resolving $\lambda$ into electron spin}\label{subsec:origin_spin}

To understand the origin of the initial decrease followed by a sudden increase in $\lambda$ as a function of $x$, we use our decomposition of $\lambda$ into the majority-spin part $\lambda^{\uparrow}$ and the minority-spin part $\lambda^{\downarrow}$, as defined in Sec~\ref{sec:methods}. We show in Fig.~\ref{fig:lambda} the $\lambda^{\uparrow}$ (red line) and $\lambda^{\downarrow}$ (blue line) as a function of the alloy composition $x$. We also show the numerical values in Table~\ref{tab:results}. The majority spin $\lambda^{\uparrow}$ is dominant over the minority spin $\lambda^{\downarrow}$ in pure Fe ($x=0$).  However, the previously dominant $\lambda^{\uparrow}$ drastically reduces in magnitude between $x=0$ and $x=0.25$, and remains small (around 0.02) until $x=0.75$. The minority $\lambda^{\downarrow}$ has the opposite behavior.  Although $\lambda^{\downarrow}$ was small in Fe ($x=0$) it increases about 12-fold as concentration $x$ of Co increases from $x=0$ to $x=0.75$. The cross-over from $\lambda^{\uparrow} > \lambda^{\downarrow}$ to $\lambda^{\uparrow} < \lambda^{\downarrow}$ occurs already around $x = 0.125$.  

Therefore, the dependence of $\lambda$ on $x$ in Fe$_{1-x}$Co$_x$ is due to the different behavior of the majority-spin and minority-spin channels. This behavior is in a clear contrast to the nonmagnetic alloys discussed in Sec.~\ref{sec:elph_alloys}, where by definition $\lambda^{\uparrow} = \lambda^{\downarrow}$.  Therefore, trivially, any nonlinearity in $\lambda(x)$ for a nonmagnetic alloy cannot come from the different dependence of $\lambda^{\uparrow} (x)$ compared to $\lambda^{\downarrow}(x)$. 

\subsection{Resolving $\lambda$ into electronic density of states and electron-phonon matrix elements}\label{subsec:origin_g_dos}

Next, we discuss the origin of the composition dependence of $\lambda^{\uparrow}$ and $\lambda^{\downarrow}$. We use Eq.~\ref{eq:lam_simple_spin} to decompose $\lambda^{\sigma}$ into a product of $\frac{\left(N^{\sigma}_{\rm F}\right)^2}{ N_{\rm F}^{\uparrow} + N_{\rm F}^{\downarrow} }$ and $\langle g_{\sigma}^2 \rangle$. We begin with an analysis of the composition dependence of the spin-resolved quantity $\frac{(N^{\sigma}_{\rm F})^2}{ N_{\rm F}^{\uparrow} + N_{\rm F}^{\downarrow} }$, as shown in the top panel of Fig.~\ref{fig:expl2} (for $N^{\sigma}_{\rm F}$ and $N_{\rm F}^{\uparrow} + N_{\rm F}^{\downarrow}$, see Table~\ref{tab:results}). We observe that $\frac{(N^{\sigma}_{\rm F})^2}{ N_{\rm F}^{\uparrow} + N_{\rm F}^{\downarrow} }$ qualitatively tracks the dependence of $\lambda^\sigma$ on composition ($x$). That is, at small $x$ the $\frac{\left(N^{\uparrow}_{\rm F}\right)^2}{ N_{\rm F}^{\uparrow} + N_{\rm F}^{\downarrow} }$ is larger than $\frac{\left(N^{\downarrow}_{\rm F} \right)^2}{ N_{\rm F}^{\uparrow} + N_{\rm F}^{\downarrow} }$ but with increasing $x$ they switch. Clearly, a large part of the dependence of $\lambda^{\sigma}$ on the composition ($x$) comes from the density of states. However, the matrix elements also play a role in the dependence of $\lambda^\uparrow$ on $x$. The bottom panel of Fig.~\ref{fig:expl2} shows the majority $\langle g_{\uparrow}^2 \rangle$ (red line) and the minority $\langle g_{\downarrow}^2 \rangle$ (blue line) as a function of $x$.  The same data are also included in Table~\ref{tab:results}.  As can be seen in the figure and the table, the average electron-phonon matrix elements depend on the Co concentration, $x$. We find that $\langle g_{\uparrow}^2 \rangle$ increases by nearly a factor of 3 from $x=0$ to $x=0.75$. Despite the increase in $\langle g_{\uparrow}^2 \rangle$, the 25-fold decrease in $\frac{\left(N^{\uparrow}_{\rm F}\right)^2}{ N_{\rm F}^{\uparrow} + N_{\rm F}^{\downarrow} }$ from $x=0$ to $x=0.75$ drives the reduction in $\lambda^\uparrow$ with $x$. On the other hand, the minority spin $\langle g_{\downarrow}^2 \rangle$ is roughly constant with $x$, the difference between the largest and smallest $\langle g_{\downarrow}^2 \rangle$ is around 20\%. We also note that for all $x$ the minority spin $\langle g_{\downarrow}^2 \rangle$ is larger than the majority spin $\langle g_{\uparrow}^2 \rangle$. 

\begin{figure*}[htp]
\includegraphics{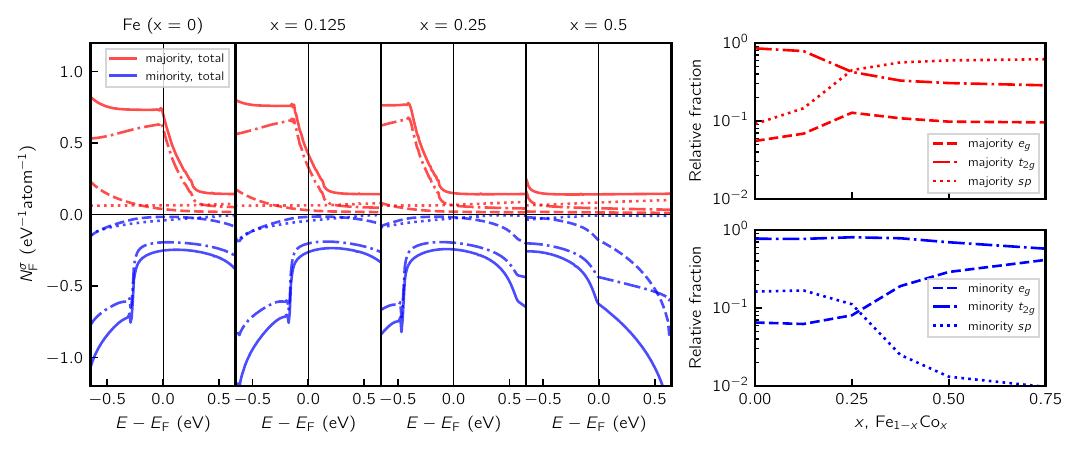}
\caption{\label{fig:orbital_char} Composition ($x$) dependence of majority (red lines) and minority (blue lines) density of states show rigid shifts in bands. Band structures at each $x$ are given in the Supplemental Materials. Projections of band onto $sp$-like (dotted lines), $t_{2 \rm g}$-like (dashed-dot lines), and $e_{\rm g}$-like (dashed lines), character are shown in the left panel. Relative fraction of each orbital character at the Fermi energy shown in the right panels.}
\end{figure*}
\begin{figure*}[htp]
\includegraphics{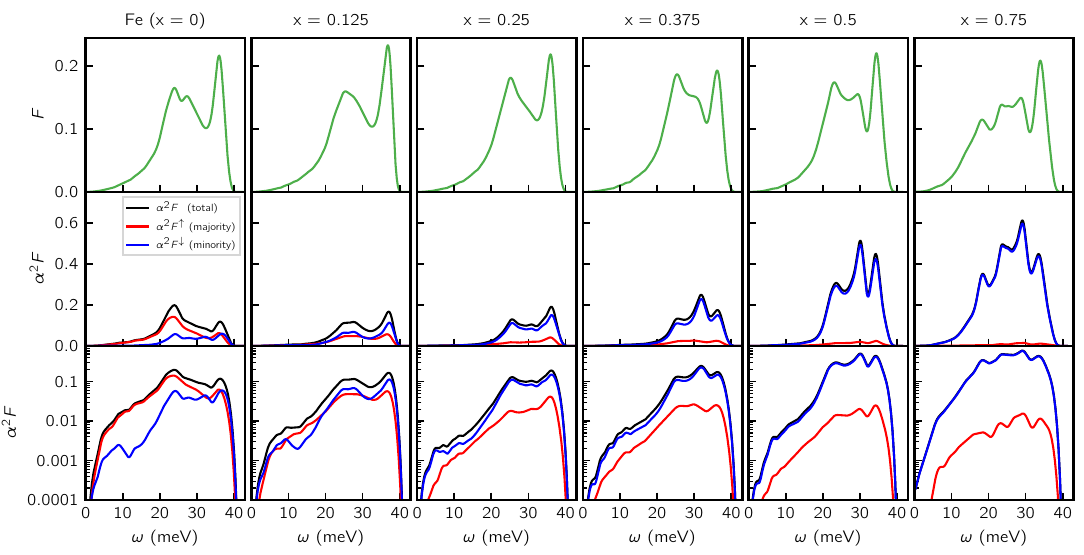}
\caption{\label{fig:expl1} Composition ($x$) dependence of the phonon density of states $F$ in the upper row and the Eliashberg spectral function for total $\alpha^2F$ (black) , majority $\alpha^2F^{\uparrow}$ (red) and minority $\alpha^2F^{\downarrow}$ (blue) in the middle row. We show in the bottom row the $\alpha^2F$ results on a logarithmic vertical scale.}
\end{figure*}

\subsection{Origin of composition dependence of electron-phonon matrix element}
Now we focus on understanding what drives the composition dependence of average electron-phonon matrix elements $\langle g_{\uparrow}^2 \rangle$ and $\langle g_{\downarrow}^2 \rangle$. The electron-phonon matrix element, as defined in Eq.~\ref{eq:g_spin}, is calculated from the electron wavefunction $\psi^{\sigma}_{nk}$ and the change in the effective electron potential due to phonon displacement $\partial_{q \nu} v$. Clearly, both $\psi^{\sigma}_{nk}$ and $\partial_{q \nu} v$ depend on the Co concentration, $x$.

First, to analyze the nature of electron wavefunction $\psi^{\sigma}_{nk}$ as a function of $x$ we decomposed electron state $\psi^{\sigma}_{nk}$ into a basis of localized Wannier functions $W^{\sigma}_i$ with a well defined orbital ($i$) and spin ($\sigma$) character,
$$
\left\vert \left\langle 
\psi^{\sigma}_{nk}
\vert  
W^{\sigma}_i
\right\rangle \right\vert^2.
$$
Importantly, the orbital character of states $W^{\sigma}_i$ is nearly independent of $x$, and therefore serves as a reference orbital against which we compare Bloch states $\psi^{\sigma}_{nk}$ for all $x$. Given $\left\vert \left\langle 
\psi^{\sigma}_{nk}
\vert  
W^{\sigma}_i
\right\rangle \right\vert^2
$ we compute the projected density of states as shown in Fig.~\ref{fig:orbital_char}.

Second, to analyze the nature of the electron potential perturbation $\partial_{q\nu} v$, we also transform it into the basis of Wannier functions, as described in Ref.~\cite{gwann}.  The transformed matrix element has the following form,
$$
g_{\sigma i j \alpha}^2 =
\left\vert \left\langle 
W^{\sigma}_i
\vert
\partial_{\alpha} v 
\vert 
W^{\sigma}_j
\right\rangle \right\vert^2.
$$
In this doubly localized representation, both the electron orbitals and the perturbing potential are now exponentially localized in space.  In particular, the change in the electron potential $\partial_{\alpha} v$ is now resulting from a displacement of a single atom, in a single (home) unit cell, in a Cartesian direction $\alpha$.  For simplicity of our following analysis, we will sum quantity $g_{\sigma i j \alpha}^2$ over all directions $\alpha$.

\subsubsection{Minority spin}

Focusing first on the minority spin, we find that for all $x$ the dominant minority electron state character at the Fermi level is $t_{2 \rm g}$.  This can be seen from the projected density of states plot shown in Fig.~\ref{fig:orbital_char}. 

Analyzing the strength of the perturbing potential $\partial v$, we find that it is nearly independent of $x$.  The maximum change with $x$ of our measure of perturbation potential strength, $\sum_{\alpha} g_{\sigma i j \alpha}^2$, is at most 15\%.  Therefore, in short, both the electron state and the perturbing potential are independent of $x$ for the minority spin.  This is consistent with our earlier observation that the average electron-phonon matrix element $\langle g_{\downarrow}^2 \rangle$ is nearly independent of $x$, as shown in Fig.~\ref{fig:expl2}. 

\subsubsection{Majority spin}

The situation with the majority spin channel is somewhat more complex than with the minority spin channel.  The dominant majority spin orbital character is $t_{2 \rm g}$ for $x$ below $0.25$.  Above $0.25$ the dominant orbital character is $sp$-like.  The orbital decomposition of the electronic states is shown in Fig.~\ref{fig:orbital_char}.  

Furthermore, the strength of the perturbing potential $\partial v$ is strongly dependent on $x$ for the majority spin channel.  As expected from the exponential localization of $g_{\sigma i j \alpha}^2$, the dominant matrix element corresponds to cases where the $i$ and $j$ orbitals are in the home unit cell.  While the $sp^3d^2$-like scattering is significantly stronger than the $t_{2 \rm g}$-like scattering, we find that the $sp^3d^2$-like scattering is less sensitive to Co concentration, $x$.  For example, going from $x = 0.25$ to $x = 0.75$ we find a nearly fourfold increase in $\sum_{\alpha} g_{\sigma i j \alpha}^2$ when $i$ and $j$ correspond to the $t_{2 \rm g}$-like orbitals and 30\% increase when $i$ and $j$ correspond to the $sp^3d^2$-like orbitals.  When $i$ or $j$ are not in the home cell, $g_{\sigma i j \alpha}^2$ is significantly smaller and even less dependent on $x$ (at most $\pm$10\%).

In short, we assign a strong increase in $\langle g_{\uparrow}^2 \rangle$ as a function of $x$ both to the change in the orbital character of the electron state at the Fermi level, as well as to the increase in the strength of the perturbing potential $\partial_{\alpha} v$.

\subsubsection{Previous work}

The importance of the orbital character of electronic states for the electron-phonon interaction strength has been discussed previously in the literature.  For example, measurements reported in Ref.~\onlinecite{PhysRevLett.92.186803} show that the matrix elements for the scattering involving $sp$ states have larger magnitude than those involving $d$ states in Ag.  Interestingly, the opposite was reported in Ref.~\onlinecite{Jamal2016} for Cu, Ag, and Au, while Ref.~\onlinecite{elec_mat_elem_pd} found a weak relationship between the magnitude of the matrix element and the orbital character. 

\subsection{Resolving $\lambda$ by phonon frequency}\label{subsec:origin_phon}

After resolving $\lambda$ into various electronic contributions, we now focus on resolving $\lambda$ by phonon frequencies. First, we analyze the phonon density of states $F(\omega)$, shown in the top row of Fig.~\ref{fig:expl1}, and the phonon dispersion (see Supplemental Materials). We see that both the phonon density of states and phonon dispersion are roughly unchanging with composition. 

To analyze which phonons contribute to $\lambda$ we computed the Eliashberg spectral function $\alpha^2F(\omega)$, as defined in Eq.~\ref{eq:lam_a2f}. The $\alpha^2F(\omega)$ keeps track of the phonon frequencies $\omega$ with a dominant contribution to $\lambda$. We show $\alpha^2F(\omega)$ for different alloy compositions, $x$, in Fig.~\ref{fig:expl1}.

In general, we find, as expected, that $\alpha^2F$ strongly varies with $x$, mirroring the strong variation discussed earlier for the integrated quantity, $\lambda$.  Focusing on pure Fe ($x=0$) we see that $\alpha^2F(\omega)$ has two peaks, one near 22~meV and another around 37~meV. A third peak, around 30~meV appears for $x$ above 0.375. The relative strength of the peaks changes as a function of $x$.  For example, at $x=0$ the dominant peak is the low-frequency one, while above $x=0.375$ the mid-frequency peak dominates, signaling a different electron-phonon scattering channels for $x$ above 0.375. Figure~\ref{fig:expl1} also shows the decomposition of $\alpha^2F$ into majority spin $\alpha^2F^{\uparrow}$ (shown in red) and minority spin $\alpha^2F^{\downarrow}$ (shown in blue).  From the spin-decomposition, we see that the origin of the low-frequency peak around 22~meV in pure Fe is from the majority-spin channel.  On the other hand, the appearance of the mid-frequency contribution above $x=0.375$ is mainly due to the minority spin channel.

\subsection{Comparison with previous works}\label{sec:prev_works}

We now compare our results with the literature results.  $\lambda$ in bcc iron was empirically estimated\cite{AllenEstimations} to be 0.9. Later calculations within the rigid muffin-tin approximation found $\lambda$ ranging around 0.5--0.9.\cite{PapaDim2014,JARLBORG198489} 

A more recent Ref.~\onlinecite{verstraete2013} reports density functional theory calculations of the electron-phonon coupling strength in pure Fe. The definition of electron-phonon coupling strength from Ref.~\onlinecite{verstraete2013} is equivalent to our non-additive $\lambda^\sigma_{\rm nadd}$ from Eq.~\ref{eq:lambda_nadd}.  The reported values of the non-additive $\lambda^\sigma_{\rm nadd}$ are 0.068 for majority spins and 0.175 for minority spins. Our calculation of the same quantity ($\lambda^\sigma_{\rm nadd}$) gives 0.222 for majority spins and 0.236 for minority spins.  Therefore, our $\lambda^\uparrow_{\rm nadd}$ is larger by about a factor of 3 while our $\lambda^\downarrow_{\rm nadd}$ is larger by a factor of 1.3 than those reported in Ref.~\onlinecite{verstraete2013}. Now we briefly analyze possible sources of the disagreement between our results and those reported in Ref.~\onlinecite{verstraete2013}. First, we recall that $\lambda^\sigma_{\rm nadd}$ is the product of $N^{\sigma}_{\rm F}$ and $\langle g_{\sigma}^2 \rangle$.  Therefore, we can investigate whether the difference with respect to Ref.~\onlinecite{verstraete2013} originates from the spin-resolved density of states $N^{\sigma}_{\rm F}$ or from the average spin-resolved matrix element, $\langle g_{\sigma}^2 \rangle$.  These quantities are given in Table~\ref{tab:results}.  Comparing the values of spin-resolved density of states we find that our $N^{\uparrow}_{\rm F}$ is 30\% larger, while $N^{\downarrow}_{\rm F}$ is 25\% smaller, than that reported in Ref.~\onlinecite{verstraete2013}.  The differences for the average matrix elements $\langle g_{\sigma}^2 \rangle$ are much larger.  Our $\langle g_{\uparrow}^2 \rangle$ is nearly 2.5 times larger while $\langle g_{\downarrow}^2 \rangle$ about 2 times smaller.  Ref.~\onlinecite{verstraete2013} reported sampling the Brillouin zone with $12^3$ $k$ points and $12^3$ $q$ points, compared to $72^3$ $k$ points and $24^3$ $q$ points that we needed to converge $\lambda$. We suspect that this difference in convergence is the main reason why the results of Ref.~\onlinecite{verstraete2013} differ from our results. 

As we described previously in Sec.~\ref{sec:measurable}, the electron-phonon energy transfer coefficient $G$ is a physically measurable quantity closely related to $\lambda$. Several values have been reported in the literature for the electron-phonon energy transfer coefficient. A semi-empirical\cite{Medvedev} model was used to estimate the electron-phonon energy transfer coefficient in pure Fe to be $20.8 \cdot 10^{17}$~W\,m$^{-3}$~K$^{-1}$, while a parameter-free model was used to obtain a value of $10.5 \cdot 10^{17}$~W\,m$^{-3}$~K$^{-1}$.\cite{Oppeneer}

The first-principles density functional theory was used in Ref.~\onlinecite{petrov2013thermal} to calculate the energy transfer coefficient as a function of the electron temperature. Extrapolation of their results to zero temperature results in an energy transfer coefficient of $7.00 \cdot 10^{17}$~W\,m$^{-3}$~K$^{-1}$.  Using Eq.~\ref{eq:G1} we get a very similar result ($7.25\cdot 10^{17}$~W\,m$^{-3}$~K$^{-1}$) for pure Fe ($x=0$). Ref.~\onlinecite{petrov2013thermal} does not report $G$ for Fe alloyed with Co. We calculate $G$ as a function of alloy composition $x$ and show the results in Table~\ref{tab:results}.

We can also compare our results for $G$ with the energy transfer coefficients obtained from the experiment. Ultrafast demagnetization measurements were used to estimate $G$ in Fe~\cite{ma14112755} (shown in Table~\ref{tab:results}) and Fe--Co alloys~\cite{Ramya2021}. Ref.~\onlinecite{Ramya2021} reports a $G$ for Fe ($x=0$) of $33 \cdot 10^{17}$~W\,m$^{-3}$~K$^{-1}$ that reaches a minimum at $x=0.25$~($11 \cdot 10^{17}$~W\,m$^{-3}$~K$^{-1}$) and then increases to $53 \cdot 10^{17}$~W\,m$^{-3}$~K$^{-1}$ at $x=0.75$. Although there is a quantitative disagreement between our results and $G$ reported in Ref.~\onlinecite{Ramya2021}, there is good qualitative agreement in the composition dependence of $G$. From $x=0$ to 0.25, we find a 2.7-fold decrease in $G$, similar to the 3.1-fold reduction found in the experiment, and from $x=0.25$ to 0.75, we find a 10-fold increase in $G$, compared to the less dramatic 5-fold increase found in the experiment. 

\section{Conclusions}\label{sec:conclusion}

The iron-cobalt alloy (Fe$_{1-x}$Co$_x$) shows strong dependence of magnetization dynamics on composition, as reported in Ref.~\onlinecite{schoen2016, Ramya2021}.  We investigated from first principles the spin-dependent strength of the electron-phonon interaction in these alloys as a function of composition ($x$).  We find a rich dependence of the electron-phonon interaction strength $\lambda$ on composition.  Analyzing separately the contributions of the majority and minority spins to $\lambda$ we find that both have strong, and opposing, variation with composition.  Interestingly, the majority component $\lambda^\uparrow$ decreases with $x$ while the minority spin increases with $x$.  We show that these compositional variations are driven by changes in both the density of states $N^\sigma_{\rm F}$ and the average electron-phonon matrix element $\langle g^2_\sigma \rangle$. 

Understanding the electron-phonon interaction strength in a ferromagnetic alloy, such as Fe$_{1-x}$Co$_x$, opens doors to a better understanding of magnetic phenomena such as Gilbert damping,\cite{Gilbert} ultrafast demagnetization,\cite{ultrafast_nickel} all-optical switching,\cite{all_optical} and spin-dependent transport.\cite{spin_peltier,spin_heat_acc,GMTR_nat,seebeck_eph} Our work quantifies the extent to which adjusting the composition of ferromagnetic alloys, such as Fe--Co, can tune the aforementioned dynamic magnetic properties.

\section{Acknowledgements}
This work was supported by the NSF DMR-1848074 grant.
\appendix

\bibliography{pap}
\end{document}